\documentclass{emulateapj}
\usepackage{apjfonts}

\begin{document}

\title{The WFC3 Galactic Bulge Treasury Program: \\
A First Look at Resolved Stellar Population Tools}

\author{
Thomas M. Brown\altaffilmark{1}, 
Kailash Sahu\altaffilmark{1}, 
Manuela Zoccali\altaffilmark{2}, 
Alvio Renzini\altaffilmark{3}, 
Henry C. Ferguson\altaffilmark{1}, 
Jay Anderson\altaffilmark{1} 
Ed Smith\altaffilmark{1}, 
Howard E. Bond\altaffilmark{1}, 
Dante Minniti\altaffilmark{2,4}, 
Jeff A. Valenti\altaffilmark{1}, 
Stefano Casertano\altaffilmark{1}, 
Mario Livio\altaffilmark{1}, 
Nino Panagia\altaffilmark{1}, 
Don A. VandenBerg\altaffilmark{5}, 
Elena Valenti\altaffilmark{6}, 
}

\altaffiltext{1}{Space Telescope Science Institute, 3700 San Martin Drive,
Baltimore, MD 21218;  tbrown@stsci.edu, ksahu@stsci.edu, 
ferguson@stsci.edu, jayander@stsci.edu, edsmith@stsci.edu, bond@stsci.edu, 
valenti@stsci.edu, stefano@stsci.edu, mlivio@stsci.edu, panagia@stsci.edu}

\altaffiltext{2}{P. Universidad Cat$\acute{\rm o}$lica de Chile, 
Departmento de Astronom$\acute{\rm i}$a y Astrof$\acute{\rm i}$sica, 
Casilla 306, Santiago 22, Chile; mzoccali@astro.puc.cl, dante@astro.puc.cl}

\altaffiltext{3}{Osservatorio Astronomico, Vicolo Dell'Osservatorio 5, 
I-35122 Padova, Italy; alvio.renzini@oapd.inaf.it}

\altaffiltext{4}{Vatican Observatory, V-00120 Vatican City State, Italy}

\altaffiltext{5}{Department of Physics and Astronomy, 
University of Victoria, P.O. Box 3055, Victoria, BC, V8W 3P6, Canada; 
davb@uvvm.uvic.ca}

\altaffiltext{6}{European Southern Observatory, Alonso de Cordova 3107,
Vitacura, Santiago, Chile; evalenti@eso.org}

\submitted{To appear in The Astronomical Journal}

\begin{abstract}
When the Wide Field Camera 3 (WFC3) is installed on the {\it Hubble
Space Telescope (HST)}, the community will have access to powerful new
capabilities for investigating resolved stellar populations.  The
WFC3 Galactic Bulge Treasury program will obtain deep imaging in
five photometric bands on four low-extinction fields. These data will
have no proprietary period, and will enable a variety of science
investigations not possible with previous data sets. To aid in planning
for the use of these data and for future observing proposals, we provide 
an introduction to the Treasury program, its photometric system, and the
associated calibration effort.

The observing strategy is based upon a new photometric system
employing five WFC3 bands spanning the UV, optical, and near-infrared:
F390W, F555W, F814W, F110W, and F160W (analogous but not identical to
the ground-based filters Washington $C$, $V$, $I$, $J$, and $H$).
With these bands, one can construct reddening-free indices of
temperature and metallicity.  Using this photometric system, the
program will target six fields in well-studied star clusters, spanning
a wide range of metallicity, and four fields in low-extinction windows
of the Galactic bulge.  The cluster data serve to calibrate the
reddening-free indices, provide empirical population templates, and
correct the transformation of theoretical isochrone libraries into the
WFC3 photometric system.  The bulge data will shed light on the bulge
formation history, and will also serve as empirical population
templates for other studies. One of the fields includes 12 candidate hosts of
extrasolar planets.

Color-magnitude diagrams (CMDs) are the most popular tool for
analyzing resolved stellar populations. However, due to degeneracies
among temperature, metallicity, and reddening in traditional CMDs,
it can be difficult to draw robust conclusions from the data.  The
five-band system used for the bulge Treasury observations will provide
reddening-free indices that are roughly orthogonal in temperature and
metallicity, and we argue that model fitting in an index-index diagram
will make better use of the information than fitting separate CMDs.  We
provide some results from simulations to show the expected quality of
the data and their potential for differentiating between different
star-formation histories.

\end{abstract}


\section{Introduction}

The Wide Field Camera 3 (WFC3) is scheduled to be installed on the
{\it Hubble Space Telescope (HST)} during the next servicing mission.
Although the camera is intended as a general-purpose instrument, it
offers particularly powerful new capabilities for studying resolved
stellar populations.  The camera will provide wide-field
high-resolution imaging with continuous spectral coverage from the
ultraviolet into the near-infrared through a choice of 77 filters and
3 grisms.  Because it will be the first {\it HST} instrument to
provide wide-field imaging in the near-infrared, an obvious target for
WFC3 is the Galactic bulge, given its high reddening and crowding.  We
describe here an observing technique that employs five WFC3 bands to
create reddening-free indices of temperature and metallicity that can be
used to explore a highly reddened and complex stellar population, 
such as the bulge.

The tools and techniques needed for a bulge investigation should be
useful to a wider range of resolved stellar population studies, and a
large database of bulge stars with reliable parameters (temperature,
metallicity, proper motion, etc.) can enable additional observing
programs directed specifically at the bulge, from both ground and
space.  These factors drove the creation of a bulge Treasury program
(GO-11664), where all of the tools and products would be
non-proprietary.  The primary purpose of this paper is to demonstrate
the general utility of the program's population tools (\S2), which include
photometry of six star clusters that can be used as empirical population
templates (\S2.3).  In
addition, this paper summarizes the scientific goals and data products
specific to the Galactic bulge (\S3).  That description of the bulge
investigation also motivates the development of the tools, provides
concrete examples of how they can be used in stellar population
studies, and enables future bulge studies.  An early look at the
program and its products allows them to be considered in the planning
of observing proposals for upcoming {\it HST} cycles.

\section{A New Photometric System for HST}

\subsection{Reddening-Free Indices}

Using techniques similar to those developed decades ago (e.g., Johnson
\& Morgan 1953; Str$\ddot{\rm o}$mgren 1966), we have constructed
reddening-free indices of $T_{\rm eff}$ and metallicity combining five
WFC3 filters (Figure 1): F390W, F555W, F814W, F110W, and F160W.  For
simplicity, these bands will hereafter be designated by names
reflecting their ground-based analogs (Washington $C$, $V$, $I$, $J$,
and $H$), and all of the magnitudes will be relative to Vega.
Although the focus here is on the specific bandpasses of the WFC3,
these photometric methods could in principle employ
ground-based photometry in similar bandpasses; however, we note that the
calibration of the indices for WFC3 will not be exactly the same as
that for other observatories.

The usual practice for deriving a star formation history from such
photometry would involve the construction of color-magnitude diagrams
(CMDs).  In a complex stellar population hosting a mix of ages and
metallicities, this practice can provide the distributions in age and
metallicity in a statistical sense.  However, these five bandpasses go
beyond such population fits to enable accurate measurements of $T_{\rm
  eff}$ and metallicity for {\it individual} stars.  Although
originating much earlier (Str$\ddot{\rm o}$mgren 1966 and references
therein), the methodology is explained well by Mihalas \& Binney
(1981), and briefly summarized here.  A reddening-free index combining
three filters takes the form of
\begin{equation}
[c] = (m_1 - m_2) - (m_2 - m_3) \times \frac{E(m_1 - m_2)}{E(m_2-m_3)}
\end{equation}
where $E(m_1 - m_2)$ is the excess in $(m_1 - m_2)$ color due to
extinction.  Insensitivity to extinction requires that the ratio
$E(m_1 - m_2) / E(m_2-m_3)$ be approximately constant with changes in
stellar parameters.  The choice of bands depends on the stellar
parameter one is trying to constrain; in broad terms, bands spanning
the optical to near-IR can constrain temperature from the peak and
redward slope of the spectrum, while bands in the UV and optical can
constrain the metallicity by exploiting UV absorption features.
Guided by these facts, one can construct the following reddening-free
indices of temperature and metallicity:
\begin{equation}
[t] = (V - J) - 5.75 \times (J - H) 
\end{equation}
and
\begin{equation}
[m] = (C - V) - 0.90 \times (V - I).
\end{equation}
The coefficients come from an analysis of stellar isochrones
(VandenBerg et al.\ 2006) folded through a library of synthetic
spectra (Castelli \& Kurucz 2003) and the response functions for each
WFC3 filter (Figure 1), under different levels of assumed reddening
(Fitzpatrick 1999).  In Figure 2, we show the variation of these
coefficients with surface gravity and effective temperature.

In practice, the coefficients, the relationship between [m] and
[Fe/H], and the relationship between [t] and $T_{\rm eff}$ should be
calibrated using observations of well-studied star clusters, because
the uncertainties are similar to those commonly found in the
derivation of isochrone transformations.  For example, when Brown et
al.\ (2005) derived the transformation of theoretical isochrones into
the bandpasses of the Advanced Camera for Surveys (ACS), they folded
the Castelli \& Kurucz (2003) spectra through the ACS response
functions and found that an empirical color correction was required to
align the transformed isochrones with the observed CMDs of star
clusters.  The color of the red giant branch (RGB) at a given
luminosity is a reddening-dependent indicator of metallicity, but the
empirical correction to the transformation is required to use the RGB
color in this manner.  It was unclear how much of this correction was
driven by uncertainties in the isochrones themselves, the synthetic
spectra, or the response functions, but there will likely be similar
issues with the transformation into WFC3 bandpasses.

Figures 3 and 4 demonstrate how these indices can be used to estimate
effective temperature and metallicity even in the presence of
significant reddening spreads and/or uncertainties.  These figures
split the population into dwarfs and giants for two reasons.  The
first reason is that dwarfs and giants overlap in a plane defined by
temperature and metallicity, in contrast to a plane defined by
temperature and luminosity (i.e., a CMD).  Second, the reddening-free
indices [m] and [t] are not completely insensitive to surface gravity;
in particular, the relation between [m] and [Fe/H] is significantly
distinct for dwarfs and giants.  In general, an uncertainty of
$\sim$0.1~mag in each index provides uncertainties of $\lesssim$300~K
in effective temperature and 0.2 dex in [Fe/H].  The high gravity
boundary in Figure 3 and the low gravity boundary in Figure 4 excludes
stars below $T_{\rm eff} \lesssim 4000$~K, where the indices vary
strongly with stellar parameters and the synthetic spectra are most
prone to systematic errors; the indices for such cool stars will
likely require significant calibration corrections using observations
of stellar clusters.  Because the coefficients in these indices are
slowly varying functions of surface gravity (Figure 2), the estimates
of effective temperature and metallicity are not particularly
sensitive to distance uncertainties.  One would use magnitude as a
proxy for surface gravity when translating [m] to [Fe/H] and [t] to
effective temperature (i.e., one can use position in a CMD to
designate a star as a dwarf or giant and use the indices
appropriately).  In the bulge Treasury program, these indices
will yield reliable estimates of metallicity and temperature for
several tens of thousands of stars in a given WFC3 field, and do so in
four fields; this will increase the number of bulge stars with such
measurements by over two orders of magnitude.  Although these indices
are not direct age indicators, the temperature distribution in a
resolved stellar population is a strong function of age.  In the
simulations below, we demonstrate how these diagrams can thus
constrain the age distribution.

\begin{figure}[t]
\epsscale{1.1}
\plotone{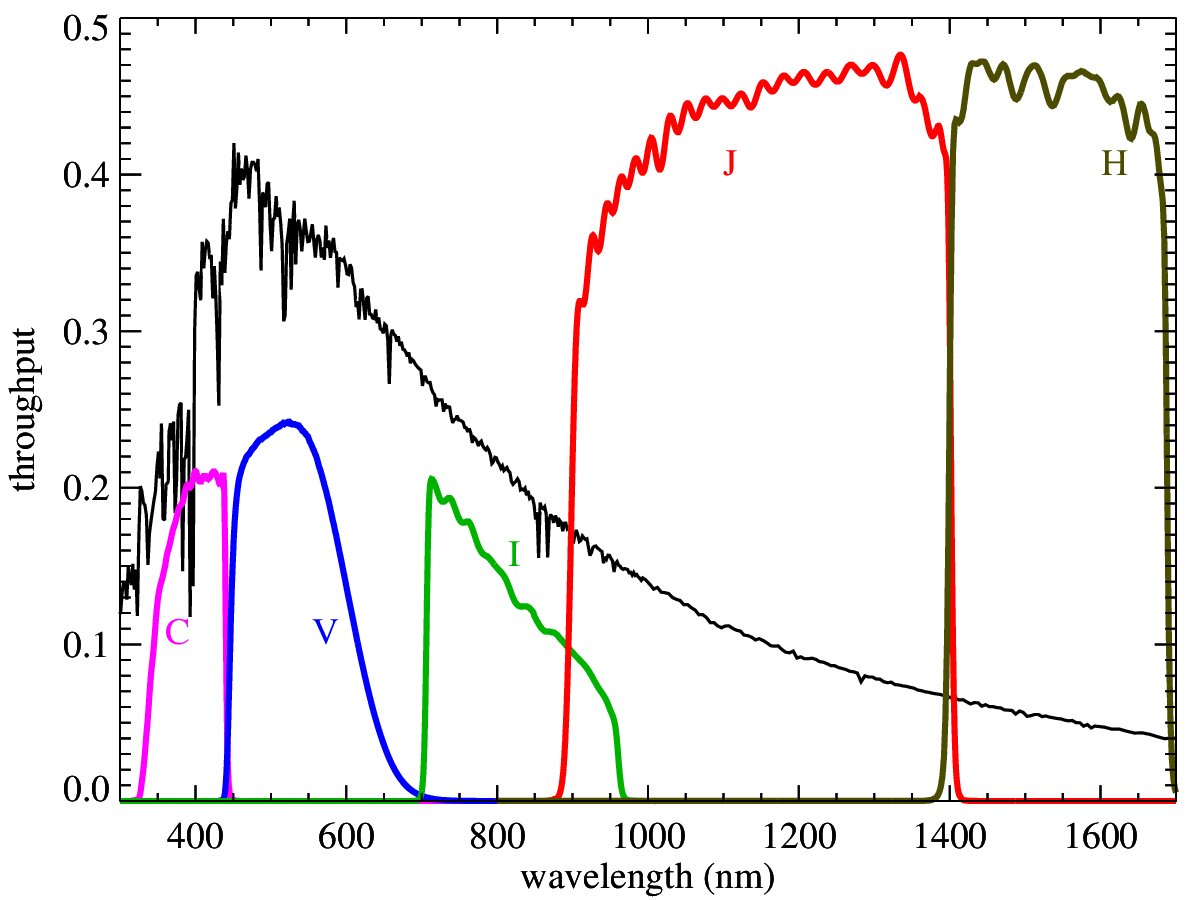}
\epsscale{1.1}
\caption{The system throughputs ({\it colored curves}) of the bandpasses
  employed in the bulge Treasury program: F390W (Washington $C$), F555W
  ($V$), F814W ($I$), F110W (wide $J$), and F160W (broad $H$).  A
  solar analog spectrum (Castelli \& Kurucz 2003; {\it black curve})
  is shown for reference.}
\end{figure}

\begin{figure}[t]
\epsscale{1.1}
\plotone{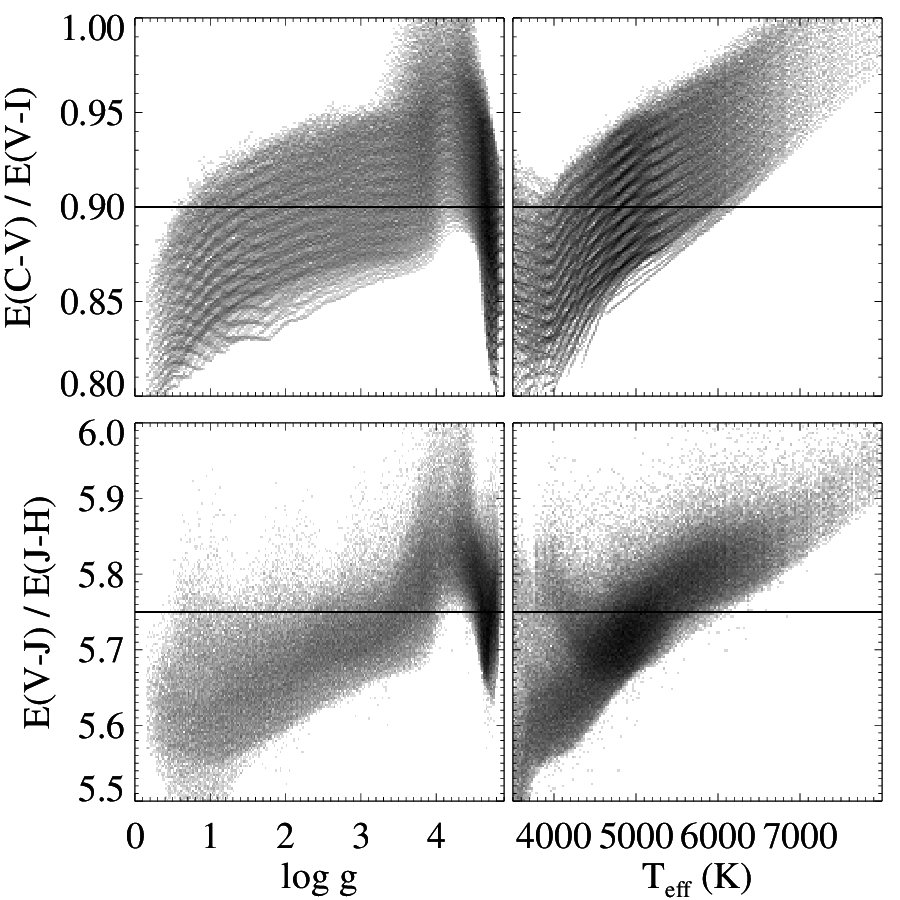}
\epsscale{1.1}
\caption{Color excess ratios as a function of stellar parameters,
  shown for isochrones ({\it grey shading}) spanning a range of age
  (4--14~Gyr), metallicity ($-2.3 \leq$~[Fe/H]~$\leq 0.5$), and
  reddening ($0 \leq A_V \leq 3.1$~mag).  The values chosen for the
  coefficients in the reddening-free indices [m] and [t] are
  highlighted ({\it black lines})}.
\end{figure}

\begin{figure}[t]
\epsscale{1.1}
\plotone{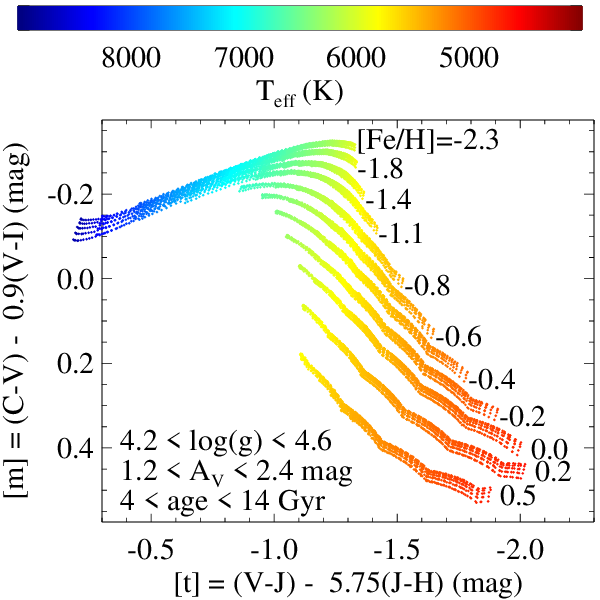}
\epsscale{1.1}
\caption{The distribution in the reddening-free indices [t]
  (temperature) and [m] (metallicity) for dwarf stars in a resolved stellar
  population spanning a wide range of age, extinction, and
  metallicity.  Each band of points is labeled with the value of
  [Fe/H], while the colors indicate $T_{\rm eff}$.  The $T_{\rm eff}$
  distribution at a given [Fe/H] is a strong function of age.  The
  small spread in the band at each value of [Fe/H] comes from the fact
  that the indices are not perfectly insensitive to reddening.
  Nevertheless, this five-band system will provide sensitive tools for
  breaking the degeneracies between [Fe/H], $T_{\rm eff}$, and
  extinction. Although analogous filters on the ground could be
  used to construct a similar diagram, it is worth noting that the
  calibration and coefficients would be distinct.  Note that [t] becomes
  more positive with increasing temperature, opposite to colors such as
  $B-V$.}
\end{figure}

\begin{figure}[t]
\epsscale{1.1}
\plotone{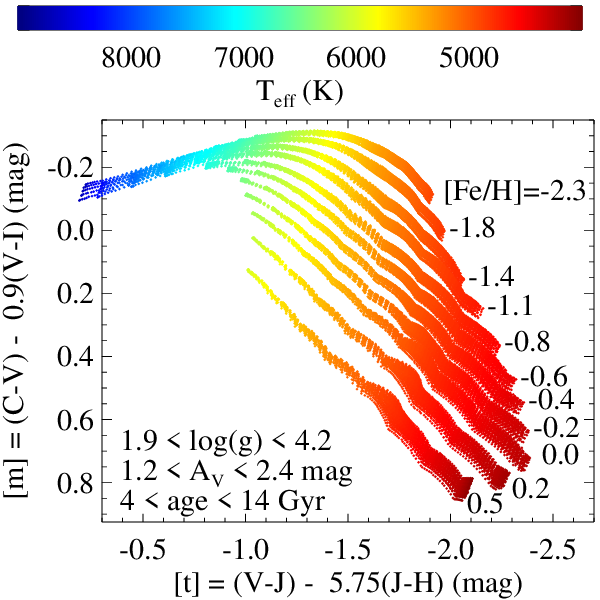}
\epsscale{1.1}
\caption{
The same as in Figure 3, but for giant stars.}
\end{figure}

To find a photometric system that would be most useful for a broad range of
stellar populations work, we folded the Victoria isochrone library
(VandenBerg et al.\ 2006) through the entire WFC3 filter set, using
the IRAF synphot package and the Castelli \& Kurucz (2003) library of
synthetic spectra.  The search used five criteria: the filters must
provide strong constraints on $T_{\rm eff}$ and [Fe/H], they must
provide such constraints despite significant distance and reddening
uncertainties (enabling use in both clusters and field populations),
the filters should be as broad as feasible (for observing efficiency),
the number of filters should be minimized (for observing efficiency),
and the bands should be as close as possible to standard photometric
systems (to enable comparisons with both archival and future
research).  We explored many alternatives (e.g., narrower filters,
alternative UV bands), but found that these were very inefficient
for observing old (cool) stellar populations and/or highly reddened
environments, and offered little improvement in the determination of
[Fe/H] and $T_{\rm eff}$.  For example, one can construct a
metallicity index that uses F336W (similar to $U$) instead of F390W.
To achieve the same [Fe/H] accuracy with this alternative index as
that achieved with the nominal [m] index above, the requirement on
signal-to-noise in the F336W photometry is only half of that 
in the F390W photometry. However, for highly-reddened ($A_V >
1.5$~mag) stars at cool temperatures ($T_{\rm eff} \lesssim$~4500~K),
the observing time required to obtain such F336W photometry is over
ten times longer than that needed to obtain the requisite F390W
photometry.

The filters most commonly used for stellar populations work on the
ACS were the F606W and F814W.  While these filters are
extremely efficient for observing faint stars, they are broad and not
well-separated in wavelength.  The WFC3 filters described here are not as
efficient as the ACS ones, but they yield much stronger constraints on
stellar parameters -- breaking degeneracies between age, [Fe/H],
and reddening (Figure 5). In their study of the stellar
populations of Andromeda, Brown et al.\ (2006) explored the various
systematic uncertainties pertinent to their ACS CMD analyses, and
found that relatively small extinction uncertainties translated into
relatively large age and [Fe/H] uncertainties.

\begin{figure*}[t]
\epsscale{1.1}
\plotone{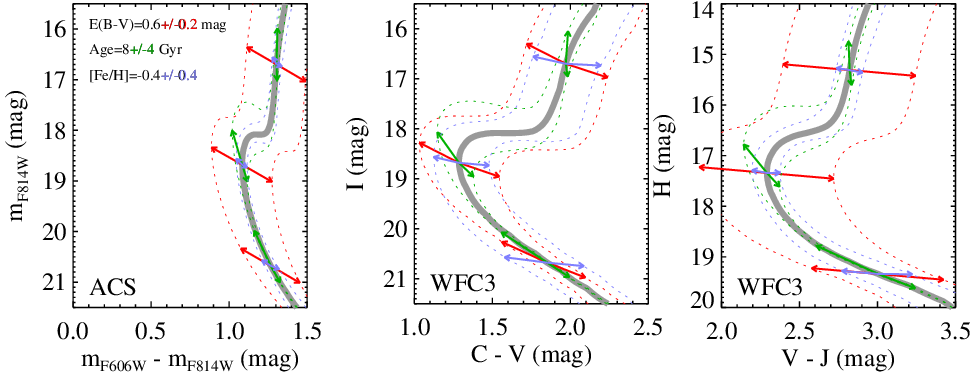}
\epsscale{1.1}
\caption{An isochrone for an 8~Gyr population with [Fe/H]~=~$-0.4$ at
  7.2 kpc with $E(B-V)=0.64$~mag ({\it grey curve}).  Three positions
  are marked with arrows: the MSTO (bluest point on
  the main sequence), a point 2~mag brighter (on the RGB), and a point
  2~mag fainter (on the main sequence).  At each position, the colored
  arrows indicate how these positions would move in CMD space for
  changes in age ({\it green}), [Fe/H] ({\it blue}), and extinction
  ({\it red}).  Dotted lines indicate how the entire isochrone would
  move.  The RGB has little sensitivity to age, while the main
  sequence (below the MSTO) has no sensitivity to age, so the age
  shifts ({\it green arrows}) at these points are simply due to the
  fact that these points are defined relative to the luminosity of the
  MSTO, which is very sensitive to age. The left
  panel shows a CMD in the ACS F606W and F814W bands, while the middle
  and right panels show CMDs constructed from the WFC3 filters,
  with the same stretch on the axes.  The WFC3 CMDs provide far
  better constraints on stellar parameters and break the degeneracies
  present in the ACS CMD.  To recover the star-formation history in
  the bulge, one would fit the distributions in all five WFC3 filters
  simultaneously (not just the CMDs shown here).}
\end{figure*}

A variety of CMDs can be constructed from these WFC3 filters, and they
can all be fit simultaneously to yield the star-formation history in a
stellar population (e.g., Figure 5).  In the case of a simple stellar
population, such fits use two fundamental
clocks present in a CMD: the luminosity difference between the main
sequence turnoff (MSTO) and horizontal branch (Sandage 1982; Iben \&
Renzini 1984), and the color difference between the MSTO and the red
giant branch (VandenBerg et al.\ 1990).  At fixed composition, the
horizontal branch luminosity and red giant branch color are both
relatively insensitive to age, while the MSTO becomes fainter and
redder at increasing age.  In the case of a complex stellar population
such as a galaxy, CMD fitting (by Maximum Likelihood or
$\chi^2$ statistics) constrains the full two-dimensional distribution
of age and [Fe/H] in a population, so that one can avoid assuming a
single age-metallicity relationship (e.g., Dolphin 2002; Harris \& Zaritsky 
2001; Gallart et al.\ 1999; Holtzman et al. 1999; Brown et al.\ 2006;
Cole et al.\ 2007).
Such age-metallicity distributions can provide a definitive check on
the chemical evolution models that have been used to explore the bulge
formation history.  It is worth stressing, however, that the advantage
of the WFC3 five-band photometric system is that it can go beyond CMD
fitting to constrain the $T_{\rm eff}$ and [Fe/H] of individual stars.
The ability to measure these parameters independent of reddening can
provide stronger constraints on the star formation history, as
demonstrated below.

\subsection{Simulated Populations}

Figure 6 shows simulations of the planned bulge observations that
demonstrate the sensitivity of this five-band photometry to the
star-formation history.  We simulated WFC3 images for each filter with
realistic exposure times, backgrounds, noise, and crowding, and then
blindly recovered the photometry using the DAOPHOT software of Stetson
(1987).  Despite the large spreads in assumed distance and extinction,
an old population that formed over a period of several Gyr ({\it
  middle panels}) is easily distinguishable from one that formed in a
nearly coeval burst ({\it bottom panels}).  This particular simulation
assumed the properties of the ``SWEEPS'' bulge field observed by Sahu et
al. (2006) with the ACS, but the formation history will be
distinguishable in each of the bulge fields being observed with WFC3.
We have conservatively assumed the Dwek et al.\ (1995) bulge model to
calculate the line-of-sight distance spread, and the small variation
in this spread from field to field.  Measurements of the red clump
luminosity within the SWEEPS field (Clarkson et al.\ 2008) demonstrate the
spread is actually somewhat smaller ($\sim$0.17~mag instead of
$\sim$0.24~mag), while ground-based surveys verify that the variation
in spread is small ($\lesssim$0.05~mag) from field to field
(Rattenbury et al.\ 2007).

\begin{figure*}[t]
\epsscale{1.1}
\plotone{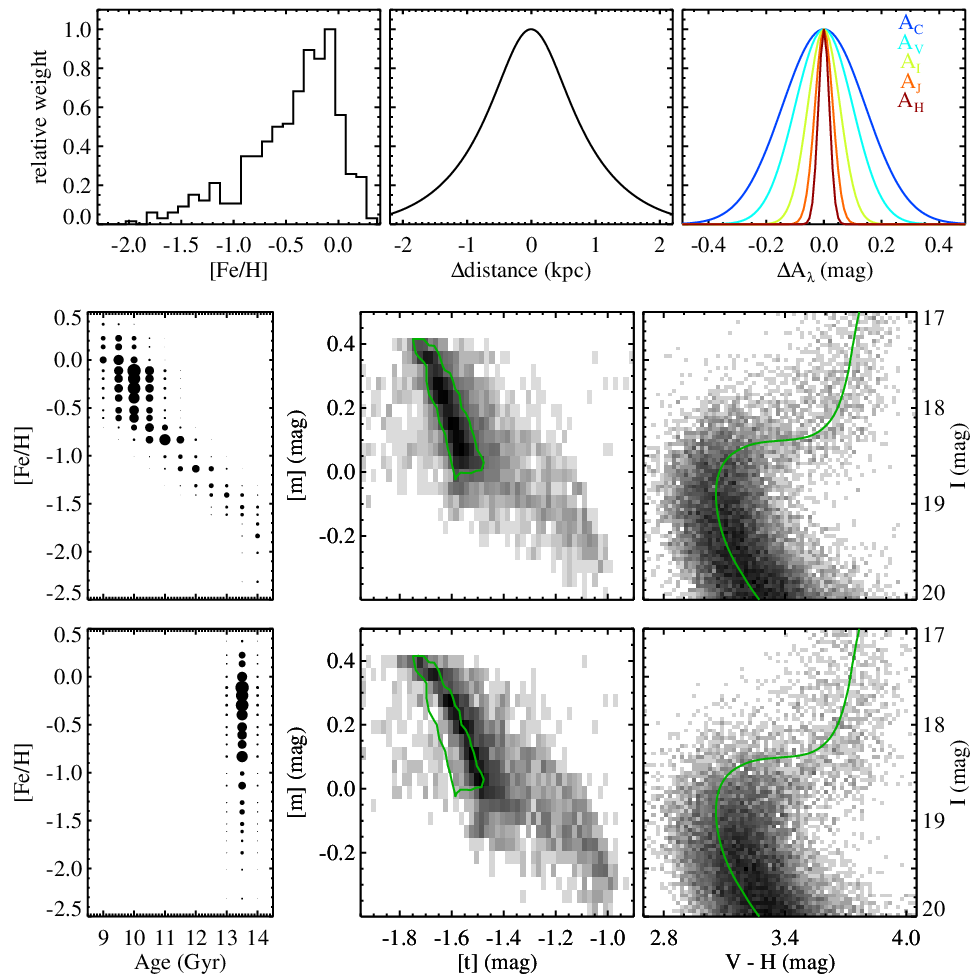}
\epsscale{1.1}
\caption{Simulations demonstrating how WFC3 bulge observations could
  distinguish between bulge formation scenarios.  {\it Top row:}
  assumptions common to each simulation -- [Fe/H] distribution
  (Zoccali et al.\ 2003), distance spread (Dwek et al.\ 1995), and
  extinction spread.  {\it Middle row:} a simulated dataset arising
  from observations of a population with an extended star formation
  history, with the population used as input ({\it left panel}; area
  of filled circles proportional to the weight at the given age and
  [Fe/H]), and two diagrams constructed from the resulting photometry:
  [m] vs.\ [t] ({\it middle panel}; for stars at $18 \leq I \leq
  19$~mag), and $I$ vs.\ $V - H$ CMD ({\it right panel}).  {\it Bottom
    row:} a simulated dataset arising from observations of an ancient
  population that formed in a nearly coeval burst, with the population
  used as input ({\it left panel}), the [m] vs.\ [t] diagram ({\it
    middle panel}; $18 \leq I \leq 19$~mag) and $I$ vs $V - H$ CMD
  ({\it right panel}).  In the [m] vs.\ [t] diagrams, the same contour
  ({\it green curve}) for the simulation in the middle row is shown
  for reference.  In the $I$ vs.\ $V - H$ CMDs, the same isochrone (10
  Gyr, [Fe/H]=$-0.5$; {\it green curve}) is shown for reference.  The
  [m] vs.\ [t] plots reflect the fact that the simulated populations
  have distinct temperature (and thus age) distributions but the same
  metallicity distributions.}
\end{figure*}

Although it is clear in Figure 6 that a CMD such as $I$ vs.\ $V - H$ is
sensitive to the star formation history, the use of five photometric
filters provides much better constraints than available from
traditional CMDs.  With five bands, one really has a low-resolution
spectrum of each star, and one can thus obtain better constraints on
the physical parameters of interest (luminosity, temperature, and
chemical composition), and then fit those to obtain the star formation
history.  When there are more than two photometric bands available, a
common approach to fitting the star formation history is to
simultaneously fit a series of CMDs constructed from combinations of
those bands (e.g., Harris \& Zaritsky 2004).  However, such fitting of
marginal distributions under-utilizes the information available,
because it discards some of the photometric links for each star.  The
analogous situation for two-band photometry is the fitting of two
distinct luminosity functions (which discards some information)
instead of fitting a CMD (which preserves all information).  For
example, with the five WFC3 bands here, one could construct and
simultaneously fit the following 4 CMDs: $C$ vs.\ $C-V$, $V$ vs.\ $V-I$,
$I$ vs $I-J$, and $J$ vs.\ $J-H$.  However, such a fit ignores the fact
that a given star in $C$ vs.\ $C-V$ space has a specific position in
$J$ vs.\ $J-H$ space.  That star shows up in both CMDs, but the fact
that it is the same star is not used.  In contrast, if one
simultaneously fits a series of [m] vs.\ [t] diagrams (where each
diagram is taken from a cut in magnitude), then one is preserving the
various photometric links for each star.  If a red bandpass is used
for the magnitude binning, this will minimize the uncertainties due to
extinction, while the [m] and [t] indices are already insensitive to
extinction.  The magnitude bins can be chosen to emphasize particular
stages of stellar evolution: bins spanning the main sequence (well
below the MSTO) or the RGB (well above the subgiant branch) will give
[m] vs.\ [t] diagrams that are primarily sensitive to [Fe/H], while
bins spanning the stars in the vicinity of the MSTO and subgiant
branch (such as those in Figure 6) will be sensitive to both age and [Fe/H].

Another way one can implement these reddening-free indices in a more
traditional analysis is to apply [Fe/H] cuts (using [m]) to the
catalog before constructing CMDs.  An example is shown in Figure 7.
Here, we have taken the simulated populations of Figure 6 and divided
each catalog into ``high [Fe/H]'' and ``low [Fe/H]'' bins, with a
boundary at [Fe/H]$\sim -0.8$ in the [m] vs.\ [t] plane.  Because the
primary difference between the two simulated populations is a movement
of metal-rich stars to intermediate ages, the shift in age (using the
traditional indicators such as the luminosity of the MSTO and subgiant
branch) is most apparent in the CMDs that use the [m] index to select
metal-rich stars.  Note that this technique is probably more useful
for qualitative (i.e., by eye) evaluations of the star formation
history instead of quantitative fitting.  If one divided a catalog
into metallicity bins using the [m] index, one could not simply fit
each sub-catalog with a library of isochrones matching the same
metallicity range implied by the index, because photometric scatter
will allow some stars from outside each metallicity bin to leak into
that metallicity bin.  Thus, one would still need to fit the CMDs for
each metallicity bin with an isochrone library spanning the full range
of possible metallicities, subjected to the same culling in [m]
vs.\ [t] space.  For quantitative fitting, it is cleaner to fit a
series of [m] vs.\ [t] diagrams in a series of appropriately chosen
magnitude bins.

\begin{figure*}[t]
\epsscale{1.1}
\plotone{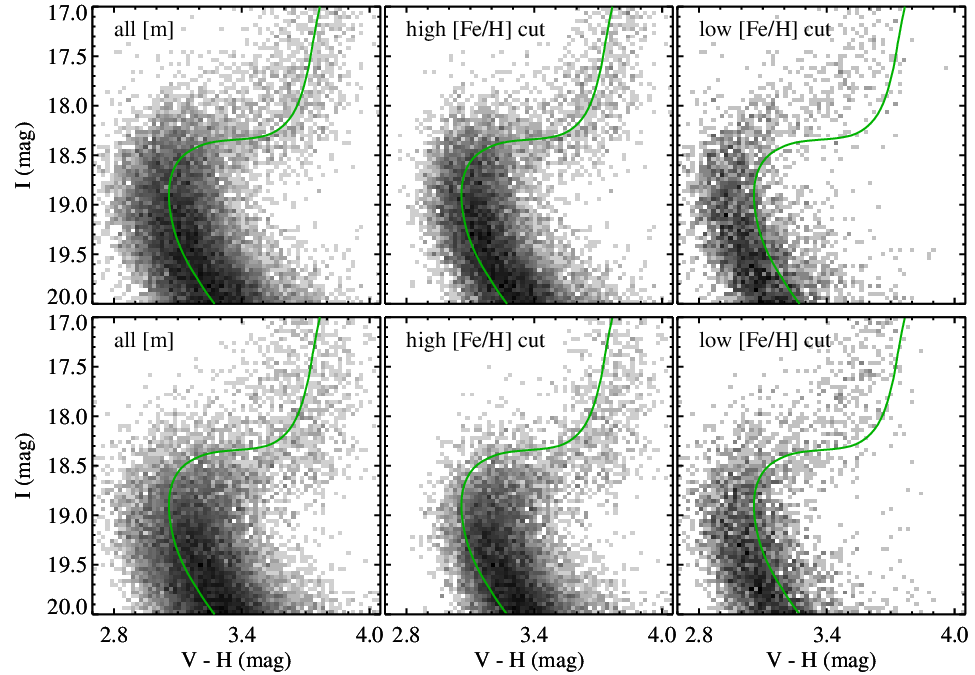}
\epsscale{1.1}
\caption{The same simulated populations shown in Figure 6 ({\it top
    panels}: extended star formation history; {\it bottom panels}:
  ancient coeval burst), but sorted into metallicity bins using cuts
  in the [m] vs.\ [t] plane: full catalog ({\it left panels}),
  metal-rich stars ({\it middle panels}), and metal-poor stars ({\it
    right panels}).  Because the primary difference between the
  simulated populations is a shift of metal-rich stars from old to
  intermediate ages (see Figure 6), the distinction between the CMDs
  is most apparent when looking at the stars at high [Fe/H] ({\it
    middle panels}).  Note that the spreads in extinction, distance, and
  age cause stars at distinct metallicities to overlap in CMD space.}
\end{figure*}

\subsection{Empirical Population Templates}

The bulge Treasury program includes observations of five Galactic
globular clusters and one open cluster (Table 1; see Brown et
al.\ 2005 and references therein for the evaluation of each cluster's
parameters).  These clusters provide empirical population templates
for direct comparison to the bulge data, enable an accurate
transformation of theoretical isochrone libraries into the WFC3
photometric system, and allow the calibration of the new
reddening-free indices of $T_{\rm eff}$ and metallicity.  Existing
synthetic spectral libraries are sufficiently accurate to demonstrate
the potential of these indices, but application of this photometric
system to real stellar populations will require direct calibration on
well-studied clusters.  These clusters were observed for a similar
purpose in a series of Large ACS programs studying the populations of
Andromeda (e.g., Brown et al.\ 2006); the cluster templates and
isochrone transformations in the ACS bandpasses were provided to the
community by Brown et al.\ (2005).  The cluster templates, isochrone
transformations, and population indices will be provided to the
community as part of the bulge Treasury program.

\begin{table}
\begin{center}
\caption{Galactic clusters}
\begin{tabular}{clllr}
\tableline
                 & $(m-M)_V$ & $E(B-V)$&        & age \\
Name             & (mag)     &  (mag)  & [Fe/H] & (Gyr)\\
\tableline
NGC~6341 (M92)   & 14.60 & 0.023& $-2.14$ & 14.5\\
NGC~6752         & 13.17 & 0.055& $-1.54$ & 14.5\\
NGC~104 (47~Tuc) & 13.27 & 0.024& $-0.70$ & 12.5\\
NGC~5927         & 15.85 & 0.42 & $-0.37$ & 12.5\\
NGC~6528         & 16.31 & 0.55 & $+0.00$ & 12.5\\
NGC~6791         & 13.50 & 0.14 & $+0.30$ &  9.0\\
\tableline
\end{tabular}
\end{center}
\end{table}

\section{Science Goals of the Treasury Program}

\subsection{The Formation History of the Bulge}

We know little about the formation of galaxy bulges, due in part to
the conflicting evidence about our own bulge, which is the one that
can be studied in greatest detail (see Minniti \& Zoccali 2007 for a
review).  From a populations perspective, the Galactic bulge looks
like a ``classical'' bulge -- i.e., similar to an old, nearly coeval
elliptical galaxy.  From a morphological perspective, the Galactic
bulge looks like a ``pseudo-bulge'' -- i.e., a peanut-shaped bulge
apparently arising from bar-driven secular processes.  As summarized
by Kormendy \& Kennicutt (2004), a range of physical processes may
contribute to bulge formation, with rapid processes in discrete events
(e.g., dissipative collapse, mergers of clouds and proto-galaxies)
dominating in the early universe, and slower secular processes
(interactions between stars, gas clouds, bars, spiral structure,
triaxial halos, etc.) dominating at later times.  In broad terms, we
see classical bulges if the earlier rapid processes are still
dominant, while pseudo-bulges are thought to arise when the later
slower processes are dominant.  This discussion highlights a tension
between observations and the framework of hierarchical galaxy
formation.  In general, semi-analytical models focus on the rapid
processes relevant to bulge formation, not on the slower secular
process; however, even these models imply that spheroid formation
occurs over billions of years -- seemingly at odds with an old, nearly
coeval spheroid (be it bulge, halo, or elliptical galaxy).

Color-magnitude diagrams of the Galactic bulge (e.g., Zoccali et
al.\ 2003; Sahu et al.\ 2006) imply an old age ($\gtrsim 10$~Gyr) with
no detected trace of an intermediate-age population, but due to the
age/metallicity/reddening degeneracies in the photometry obtained to
date, it is unclear if the star-formation history in the bulge is
somewhat extended (e.g., ages of 9--13 Gyr) or if it is restricted to
a nearly coeval burst (e.g., $13 \pm 0.2$~Gyr).  Given these
limitations, researchers have also pursued a less direct approach to
constraining the star-formation history of the bulge, by fitting
chemical evolution models to the observed metallicity distribution in
the population and the significant $\alpha$-element enhancement in
individual stars (e.g., Ferreras et al.\ 2003; Ballero et al.\ 2007).
These studies argue strongly that the bulge chemistry can be
reproduced only if the period of star formation was shorter than
1~Gyr.  Thus, both the CMD analyses (with large uncertainties) and
chemical evolution models (with indirect constraints) are consistent
with a classical bulge hosting an old, nearly coeval population.  Such
a brief burst of star formation argues against any significant
contribution from the disk or even any protracted period of
hierarchical merging.  Semi-analytic models generally imply that
bulges in galaxies like the Milky Way form over an extended period;
e.g., if one looks at Milky Way analogs in the Millennium Simulation
Project (De Lucia \& Baizot 2007; Springel et al.\ 2005), only 15\% of
the stellar mass in the bulge was present by $z \sim 1.5$.

Infrared observations clearly show that the Galactic bulge hosts a
prominent bar and that the bulge isophotes are ``boxy'' (e.g., Binney
et al.\ 1991; Blitz \& Spergel 1991; Dwek et al.\ 1995; see Figure 8).
As noted by Kormendy \& Kennicutt (2004), the only known mechanism for
producing such a structure is secular evolution of the disk (bars are
a disk phenomenon).  However, the existence of a bar does not
necessarily imply that the bulge contains young stars -- it could have
resulted from the perturbation of a pre-existing old disk.  Even so,
it is difficult to see how a bulge with a significant disk component
would be consistent with the old and nearly coeval population favored
by the chemical evolution models above.

\begin{figure}[t]
\epsscale{1.1}
\plotone{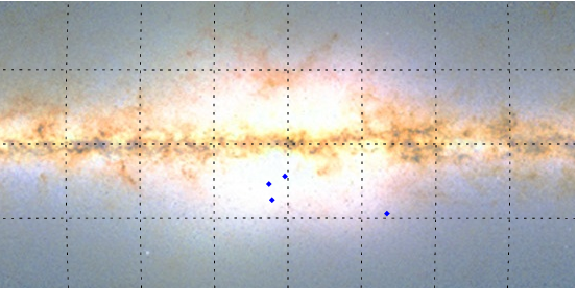}
\epsscale{1.1}
\caption{
A 2MASS image (Skrutskie et al.\ 2006)
with our bulge fields marked ({\it blue diamonds}) and a 5$^{\rm o}$
grid shown for scale ({\it dashed lines}).
}
\end{figure}

Most recently, observations of galaxies at $z\sim 2$, i.e., at a
lookback time comparable to the age of bulge stars, have opened new
opportunities for understanding bulge formation processes.  Integral
field spectroscopy of several such galaxies with $\sim$~0.2~arcsec
($\sim$1~kpc) resolution have revealed large, clumpy, rotating disks,
with high velocity dispersion ($\sim$$70-100$ km s$^{-1}$), high star
formation rate ($\sim$100~$M_\odot$~yr$^{-1}$), and very high gas
fraction (Genzel et al.\ 2006, 2008; Bouch\'e et al.\ 2007;
F\"orster-Schreiber et al.\ 2008, in prep.). Such disks are
likely to be prone to gravitational instability developing over a much
shorter timescale compared to that of typical disks at $z\sim 0$,
where instability was proposed to give rise to pseudo-bulges in
a long-lasting secular evolution of disks (Kormendy \& Kennicutt
2004). In addition to fast secular evolution of massive early disks,
migration to the center and coalescence of their giant clumps has also
been proposed for the fast formation of bulges in the early universe
(Elmegreen et al.\ 2008). These unanticipated developments have
significantly changed the traditional alternative between bulges
originating via disk evolution, hence characterized by a broad
distribution of stellar ages, and formation by massive bursts at early
cosmic times, leading to uniformly old stellar ages.  If $z\sim 2$
disks are able to make bulges within a short timescale, then the
Galactic bulge may well have formed in the same way, some 10 Gyr ago.

In addition to fast disk evolution at early times, a long lasting
secular evolution may also follow at later times, funneling younger
stars towards the bulge. If such a late time secular evolution plays a
major role in the bulge, one might expect the age distribution to
broaden as one moves from the bulge interior to its edges and
corners. Although there is evidence of a metallicity gradient, such
that metallicity decreases with increasing minor axis distance
(Zoccali et al.\ 2008; Lecureur et al.\ 2008, in prep.), the existence
of the metallicity gradient is disputed (Rich et al.\ 2007);
furthermore, the existing data give no insight into any age gradient
that might be present.  Studies of other spiral bulges in integrated
light tend to show decreasing metallicity and increasing ages with
increasing radius (e.g., Jablonka et al.\ 2007), but the statistical
and systematic uncertainties in such work are large.  

\begin{table}
\begin{center}
\caption{Bulge Fields}
\begin{tabular}{clllr}
\tableline
                & $l$      & $b$     & $A_V$   &  \\
Name            & (deg)    & (deg)   & (mag)   & Reference \\
\tableline
SWEEPS          & +1.2559  & -2.6507 & 2.0     & Sahu et al. (2006) \\
Stanek's Window & +0.2508  & -2.1531 & 2.6     & Stanek (1998) \\
Baade's Window  & +1.0624  & -3.8059 & 1.6     & Baade (1963)\\
OGLE29          & -6.7532  & -4.7195 & 1.5     & Sumi (2004) \\
\tableline
\end{tabular}
\end{center}
\end{table}

Because the bulge is largely obscured by dust, the Treasury program
will target four well-studied low-extinction windows to sample diverse
bulge fields: three near the minor axis and one in the corner of the
boxy bulge (Table 2; Figure 8).  The SWEEPS field (Sahu et al.\ 2006)
has the deepest extant optical imaging in the bulge; it was observed
with the ACS in the F606W and F814W filters, with followup to provide
proper motions (Clarkson et al.\ 2008).  Data from 2MASS (Two Micron
All Sky Survey; Skrutskie et al.\ 2006) and OGLE (Optical
Gravitational Lensing Experiment; Udalski et al.\ 2002) was used to
judge the stellar density in the other fields relative to the SWEEPS
field.  The addition of a field in Stanek's Window (Stanek 1998) will
allow a characterization of field-to-field variations in the bulge
interior.  The field in Baade's Window (Baade 1963) will provide a
minor-axis measurement closer to the edge of the boxy bulge.  Finally,
the OGLE29 field (Sumi 2004) is in one of the bulge corners.  The
OGLE29 field is also being imaged in a program using the High Acuity
Wide-field $K$-band Imaging (HAWK-I) instrument on the Very Large
Telescope (PI M. Zoccali).  The HAWK-I images will each cover an area
8$\times$ larger than a WFC3 field, searching for any trace
populations (at the level of a few percent) of young stars ($<$8~Gyr),
but without the sensitive age diagnostics possible with WFC3.  The
WFC3 program complements the HAWK-I program by obtaining the detailed
star-formation history for the population in a bulge corner using the
same photometric system employed in the interior. Proper motions of individual
stars in these fields will enable both rejection of foreground disk
stars and correlations of measured properties with bulge dynamics; as
demonstrated by Clarkson et al.\ (2008), a 1-$\sigma$ cut in proper
motion space will reduce foreground disk contamination from $\sim$10\%
to less than 1\%, yielding a nearly pure bulge population for study.

The Milky Way bulge is the only bulge where one can resolve the stars
on the old main sequence, and this will be the situation
for the foreseeable future.  As such, the Galactic bulge is under intense
scrutiny through a variety of experiments and theoretical modeling
efforts exploring the stellar populations, dynamics, dark matter, and
formation history.  The WFC3 bulge Treasury program will produce catalogs
that will be useful to such work, because its CMDs and
proper motions will distill the stars into age, metallicity, and
dynamical populations, and do so in 4 well-studied low-extinction
windows spanning a range of bulge environments.

\subsection{Jovian Planet Hosts in the Bulge}

Over 300 extrasolar planets have been discovered to date, mostly via
radial-velocity (RV) measurements, and more than 50 have been
found via photometric transit surveys. One of the most surprising results 
of these searches has been the discovery of ``hot Jupiters'' -- gas
giants with orbital periods of just a few days.  About 12 such planets
were known before the SWEEPS (Sagittarius Window Eclipsing Extrasolar
Planet Search) ACS survey.  This transit survey continuously monitored
a Galactic bulge field for 7 days, detecting 16 transiting hot Jupiter
candidates (Sahu et al.\ 2006), including a new class of
ultra-short-period planets (USPPs) orbiting with periods of
0.4--1.0~days around stars with masses less than $\sim$0.9~$M_\odot$.
Two of the brightest planet candidates were strengthened via RV
followup measurements, but the remaining candidates are too faint and
crowded for ground-based confirmation.  The SWEEPS survey was
performed in the F606W and F814W filters; while optimal for a transit
search, these filters provide inadequate constraints on the
metallicities of individual stars in the field (Figure 5), due to
degeneracies in the ACS CMD.

The bulge Treasury program will accurately characterize the effective
temperature and metallicity of 13 candidate planet hosts from SWEEPS,
including 3 USPPs.  Uncertainties of 0.2 dex in [Fe/H] and
$\lesssim$300~K in $T_{\rm eff}$ will be obtained for 12 of these
candidates, with errors twice as large for the 13$^{\rm th}$; the
remaining 3 SWEEPS candidates fall outside the WFC3 field of view.  The
program will compare the metallicity distribution for hot Jupiter hosts with
that of the general population, and do so in an environment with a
wide metallicity range ($-1.5 <$~[Fe/H]~$<+0.5$); it will also
determine if the new class of USPPs forms preferentially around
high-metallicity stars, as expected (Sahu et al.\ 2006; Ogilvie \& Lin
2007).  Metallicity measurements in the distinct bulge environment are
an important complement to extrasolar planet studies in the solar
neighborhood; the latter are mostly performed via RV measurements for
stars at $-0.5 < $~[Fe/H]~$< +0.5$, and find that planet frequency is
positively correlated with host metallicity (Fischer \& Valenti 2005).

\subsection{Metallicity Dependence of the Stellar Mass Function}

In each of the bulge fields, the photometry will yield 0.2 dex errors
in [Fe/H] for every star down to 0.5~$M_\odot$, with coarser
metallicity bins below this point (e.g., 0.4 dex errors in [Fe/H] at
0.4~$M_\odot$).  Using {\it HST,} Zoccali et al.\ (2000) measured the
bulge mass function down to 0.15~$M_\odot$ in a single bulge field;
they found the mass function was close to Salpeter above
$\sim$0.5~$M_\odot$, but more shallow at lower masses.  The bulge Treasury
program will extend this earlier work by measuring the mass function
as a function of metallicity and bulge environment, thus revealing how
the characteristic mass of star formation varies with chemistry.
Investigating the metallicity dependence of the stellar mass function
in a field population, such as the bulge, may give a better indication
of the ``initial'' mass function than studies that focus on globular
clusters.  Globular clusters generally offer clean samples at a single
age and metallicity, but the low-mass end of the mass function they
provide has been significantly distorted by dynamical processes.

\subsection{Proper Motions in the Bulge}

Proper motions of individual
stars in these fields will enable both rejection of foreground disk
stars and correlations of measured properties with bulge dynamics; as
demonstrated by Clarkson et al.\ (2008), a 1-$\sigma$ cut in proper
motion space will reduce foreground disk contamination from $\sim$10\%
to less than 1\%, yielding a nearly pure bulge population for study.
One of the four bulge fields (SWEEPS) already has proper motions with
0.3 mas yr$^{-1}$ accuracy (Clarkson et al.\ 2008) from deep
multi-epoch ACS images spanning a two-year baseline.  
The WFC3 images will be deeper than the second epoch ACS images of the
SWEEPS field, and will increase the proper motion baseline from 2
years to at least 5 years, thus significantly reducing the
uncertainties in individual proper motion measurements. For brighter
stars ($I \lesssim 24$~mag), the errors in proper motions will be
reduced by a factor of $\sim$2.5, but this increased accuracy will do
little to assist the bulge-disk separation for such stars, because the
measurement uncertainties are already significantly smaller than the
motions themselves. For fainter stars, the uncertainties in proper
motions are significant compared to their motions, driven by large astrometric
uncertainties in the second epoch.  With the deep WFC3 images, the reduction in
errors for these stars will be larger than 2.5, varying with brightness, which
will improve the bulge-disk separations.  This will be particularly
helpful in investigating the apparent discrepancy between the
theoretical isochrones and the observed CMD at these fainter
magnitudes.

For the other
three bulge fields, the Treasury program will spend three orbits to
obtain a second epoch of F814W imaging and thus proper motions to the
same accuracy.  Extant shallow ACS mosaics (using the F435W, F625W,
and F658N filters) can be used with the WFC3 images to obtain
preliminary proper motions in 2 of these fields (in Baade's Window and
Stanek's Window), but these will be hampered by differences in depth,
geometric distortion, plate scale, and bandpass.  The primary purpose
of proper motions will be to screen the foreground disk stars, but the
accuracy will be sufficient to study the bulge dynamics, complementing
existing radial velocity measurements.  The current kinematic picture
of the Galactic bulge (see Minniti \& Zoccali 2007 and references
therein) shows it to be intermediate between a rotationally-supported
system (such as the disk) and a pressure-supported system (such as the
halo), but the Treasury program will extend these measurements along
the minor axis and to a corner of the boxy bulge.

\section{Summary}

With its broad wavelength coverage, high spatial resolution, rich
filter set, and wide field of view, WFC3 brings powerful new stellar
population tools to {\it HST}.  Taking advantage of these
capabilities, the new five-band photometric system described here will
enable the construction of reddening-free indices of metallicity and
temperature.  With these indices, observers can fit the star formation
history in a population by modeling the distribution of stars in the
$T_{\rm eff}$-[Fe/H] plane, avoiding the reddening uncertainties
arising in traditional CMD fitting.  Although this system was designed
with the specific goal of understanding the formation of the Galactic
bulge, it should be useful to a wide range of stellar populations work
with WFC3.  Furthermore, because it is based upon analogs to
ground-based bandpasses, it could in principle be used in other
observatories (if properly calibrated to the exact bandpasses used).
The Treasury program will provide population templates, isochrone
transformations, and photometric indices in this system early in the
WFC3 mission.  Because the bulge science described herein is part of
the Treasury program, it will also provide accurate catalogs of
metallicity, temperature, and proper motion for hundreds of thousands
of bulge stars; these should be useful to other observational
and theoretical efforts directed at the bulge.  We encourage observers
to consider the availability of these products when planning stellar
populations proposals with WFC3 in the upcoming {\it HST} observing
cycles.

\acknowledgements

Support for proposal 11664 is provided by NASA through a grant from
the Space Telescope Science Institute, which is operated by AURA,
Inc., under NASA contract NAS 5-26555.  DM and MZ are supported by the
Center for Astrophysics FONDAP 15010003, CATA PFB-06, and Fondecyt
Regular Program \#1085278.


\begin{references}

\reference{} %
Baade, W. 1963, Evolution of Stars and Galaxies (Cambridge: Harvard
University Press), 277

\reference{} %
Ballero, S.K., Matteucci, F., Origlia, L., \& Rich, R.M. 2007, A\&A, 467, 123

\reference{} %
Binney, J., Gerhard, O.E., Stark, A.A., Bally, J., \& Uchida, K.I. 1991, 
MNRAS, 252, 210

\reference{} %
Blitz, L., \& Spergel, D.N. 1991, ApJ, 379, 631

\reference{} %
Bouch\'e, N., et al. 2007, ApJ, 671, 303

\reference{} %
Brown, T.M., et al.\ 2005, AJ, 130, 1693

\reference{} %
Brown, T.M., Smith, E., Ferguson, H.C., Rich, R.M., Guhathakurta, P., 
Renzini, A., Sweigart, A.V., \& Kimble, R.A. 2006, ApJ, 652, 323

\reference{}
Castelli, F., \& Kurucz, R.L. 2003, in IAU Symposium 210, Modeling
of Stellar Atmospheres, eds. N. Piskunov, W.W. Weiss, \& D.F. Gray, poster
A20, astro-ph/0405087 

\reference{}
Clarkson, W., et al.\ 2008, ApJ, accepted

\reference{}
Cole, A.A., et al.\ 2007, ApJ, 659, L17

\reference{}
De Lucia, G., \& Blaizot, J. 2007, MNRAS, 375, 2

\reference{}
Dwek, E., Arendt, R.G., Hauser, M.G., Kelsall, T., Lisse, C.M., Moseley, S.H.,
Silverberg, R.F., Sodroski, T.J., \& Weiland, J.L. 1995, ApJ, 445, 716

\reference{}
Dolphin, A.E. 2002, MNRAS, 332, 91

\reference{}
Elmegreen, B.G., Bournaud, F. \& Elmegreen, D.M. 2008, ApJ, in press,
   arXiv/0808.0716

\reference{}
Ferreras, I., Wyse, R.F.G., \& Silk, J. 2003, MNRAS, 345, 138

\reference{}
Fischer, D.A., \& Valenti, J. 2005, ApJ, 622, 1102

\reference{}
Fitzpatrick, E.L. 1999, PASP, 111, 63

\reference{}
Gallart, C., Freedman, W.L., Aparicio, A., 
Bertelli, G., \& Chiosi, C. 1999, AJ, 118, 2245

\reference{}
Genzel, R., et al. 2006, Nature, 442, 786

\reference{}
Genzel, R., et al. 2008, ApJ, in press, arXiv/0807.1184

\reference{}
Harris, J., \& Zaritsky, D. 2001, ApJS, 136, 25

\reference{}
Harris, J., \& Zaritsky, D. 2004, AJ, 127, 1531

\reference{}
Holtzman, J.A., et al. 1999, AJ, 118, 2262

\reference{}
Iben, I., \& Renzini, A. 1984, PhR, 105, 329

\reference{}
Johnson, H.L., \& Morgan, W.W. 1953, ApJ, 117, 313

\reference{}
Jablonka, P., Gorgas, J., \& Goudfrooij, P. 2007, A\&A, 474, 763

\reference{}
Kormendy, J., \& Kennicutt, R.C. 2004, ARAA, 42, 603

\reference{}
Mihalas, D., \& Binney, J. 1981, Galactic Astronomy (New York: W.H. Freeman
and Company), 186

\reference{}
Minniti, D., \& Zoccali, M. 2007, Galactic Bulges (San Francisco: ASP),
astro-ph/0710.3104

\reference{}
Ogilvie, G.I., \& Lin, D.N.C. 2007, ApJ, 661, 1180

\reference{}
Rattenbury, N.J., Mao, S., Debattista, V.P., Sumi, T., Gerhard, O., 
\& De Lorenzi, F. 2007, MNRAS, 378, 1165

\reference{}
Rich, R.M., Origlia, L., \& Valenti, E.\ 2007, ApJ, 665, L119

\reference{}
Sahu, K., et al.\ 2006, Nature, 443, 534

\reference{}
Sandage, A. 1982, ApJ, 252, 553

\reference{}
Skrutskie, M.F., et al.\ 2006, AJ, 131, 1163

\reference{}
Springel, V., et al.\ 2005, Nature, 435, 629

\reference{}
Stanek, K.Z. 1998, ApJL, submitted, astro-ph/9802307

\reference{}
Stetson, P. 1987, PASP, 99, 191

\reference{}
Str$\ddot{\rm o}$mgren, B. 1966, ARAA 4, 433

\reference{}
Sumi, T. 2004, MNRAS, 349, 193

\reference{}
Udalski, A., Szymanski, M., Kubiak, M., Pietrzynski, G.,
Soszynski, I., Wozniak., P., Zebrun, K., Szewczyk., O., 
\& Wyrzykowski, L. 2002, Acta Astron., 52, 217

\reference{}
VandenBerg, D.A., Bolte, M., \& Stetson, P.B. 1990, AJ, 100, 445

\reference{}
VandenBerg, D.A., Bergbusch, P.A., \& Dowler, P.D. 2006, 
ApJS, 162, 375

\reference{}
Zoccali, M., Cassisi, S., Frogel, J.A., Gould, A., Ortolani, S.,
Renzini, A., Rich, R.M., \& Stephens, A.W. 2000, ApJ, 530, 418

\reference{}
Zoccali, M., et al.\ 2008, A\&A, 486, 177

\reference{}
Zoccali, M., Renzini., A., Ortolani, S., Greggio, L., Saviane, I.,
Cassisi., S., Rejkuba, M., Barbuy, B., Rich, R.M., \& Bica, E. 
2003, A\&A, 399, 931

\end{references}
\end{document}